\documentclass{article}

\usepackage{arxiv}

\usepackage[utf8]{inputenc} 
\usepackage[T1]{fontenc}    
\usepackage{hyperref}       
\usepackage{url}            
\usepackage{booktabs}       
\usepackage{amsfonts}       
\usepackage{nicefrac}       
\usepackage{microtype}      
\usepackage{lipsum}		
\usepackage{graphicx}
\usepackage{doi}
\usepackage{float}
\usepackage{amsmath}
\usepackage{tikz}
\usepackage{caption}
\usepackage{subcaption}
\usetikzlibrary{quantikz}
\usetikzlibrary{angles,quotes}

\title{Euler-Rodrigues Parameters: A Quantum Circuit to Calculate Rigid-Body Rotations }

\author{
    Emilio Peláez \\
    The University of Chicago \\
    Chicago, Illinois, 60637, USA \\
    \texttt{epelaez@uchicago.edu} \\
    \And
    Anuranan Das \\
	Department of Electrical Engineering\\
	Indian Institute of Technology Bombay\\
    Powai, Mumbai, 400076, India\\
	\texttt{anuranan@ee.iitb.ac.in} \\

	\And
	Parmeet Singh Chani \\
	Department of Applied Physics\\
	Delhi Technological University\\
	Bawana, Delhi, 110042, India \\
	\texttt{parmeetsinghchani\_{2k19ep066}@dtu.ac.in} \\
	\And
	Daniel Sierra-Sosa \\
	Department of Computer Science \& IT \\
	Hood College \\
	401 Rosemont Ave, Frederick, MD 21701, USA \\
	\texttt{sierra-sosa@hood.edu} \\
}

\renewcommand{\headeright}{Technical Report}
\renewcommand{\undertitle}{Technical Report}
\renewcommand{\shorttitle}{\textit{arXiv} Template}

\hypersetup{
pdftitle={EULER-RODRIGUES PARAMETERS: A QUANTUM CIRCUIT TO
CALCULATE RIGID-BODY ROTATIONS},
pdfsubject={q-bio.NC, q-bio.QM},
pdfauthor={David S.~Hippocampus, Elias D.~Striatum},
pdfkeywords={Quantum Computing, Linear Algebra Trasformation, 3D Rotation},
}

\begin{document}
\maketitle

\begin{abstract}
The use of vectorial parameterization to create geometrical representations in computational models has a large number of applications. One particular application is the calculation of the 3D rotational motion of rigid bodies, that could be used for the spatial location estimation from objects. Provided the algebraic nature of this problem, it could benefit from Quantum Computing, in particular several vectors could be superposed to be transformed with a single operation, providing a quantum processing advantage.  In this article, we propose an implementation of a Quantum Computing algorithm to compute Euler-Rodrigues Parameters to model rigid body rotations to transform arbitrary functions, rotating multiple vectors in superposition. We developed this algorithm using Qiskit, taking into account the limitations imposed by the current Noisy Intermediate Scale Quantum (NISQ) devices, such as the reduced number of qubits available and the limited coherence time. 

\end{abstract}

\keywords{Quantum Computing\and Linear Algebra Transformation\and 3D Rotation}

\section{Introduction}
Quantum mechanics postulates are of algebraic nature, which means there exists an intrinsic relation between quantum computation and algebraic operations \cite{Preskill_2018}. Multiple advances in the field of quantum information processing have provided promising prospects relying on that advantage. Therefore, it has been proven that quantum computing could lead to exponential speed-up in different fields, including Principal Component Analysis (PCA) \cite{Lloyd_2014}, $k$-means clustering \cite{Lloyd_2013} and regularized Support Vector Machines (SVM) \cite{Rebentrost_2014}. Advances in NISQ devices imply the development of more diverse and meaningful applications, increasing the relevance of conducting research in this area.

One of the applications that could profit from quantum computing advantages is the vector parametrization, which relies on algebraic operations to model real-world phenomena. These parameters are commonly used in computer vision, enabling applications on engineering and applied science such as the rotational motion of rigid bodies and the estimation of 3-D rigid body transformations. Some special cases of the vector parameterization are the mathematical models including Rodrigues rotation, Cayley transform, Euler-Rodrigues equations, Gibbs, Wiener, and Milenkovic parameters \cite{Wei_2017}. All of these are characterized by a set of three parameters which behave as the Cartesian components of a geometric vector in a three-dimensional space.

Euler-Rodrigues Parameters (ERP) have received less attention in the literature than their isomorphs, quaternions, because ERP in general require more computational resources to be calculated. Taking aside the computational burden, ERP provide a simpler model interpretation and allows to obtain an invariant rotation matrix. In an invariant matrix the quantities change according to a well-known model, a transformation in linear algebra. Quantum computing could enable the calculation of these parameters given the exponential speed-up of some linear algebra operations.  
This work aims to study the feasibility of using quantum computing circuits to compute ERP in particular rigid body rotations, considering the limitations imposed by the current NISQ devices such as the reduced number of qubits available and the limited coherence time.


\section{Background and notation}

In this section, we will introduce the necessary basic concepts about Euler-Rodrigues parameters, rotations, and quantum computation.

\subsection{Rodrigues' rotation matrix}

The matrix of a proper rotation $R$ by angle $\theta$ around the axis $u = (u_x, u_y, u_z)$, a vector with $u^{2}_{x}+u^{2}_{y}+u^{2}_{z}=1$, is given by:
\begin{align}
     R={\begin{bmatrix}\cos \theta +u_{x}^{2}\left(1-\cos \theta \right)&u_{x}u_{y}\left(1-\cos \theta \right)-u_{z}\sin \theta &u_{x}u_{z}\left(1-\cos \theta \right)+u_{y}\sin \theta \\u_{y}u_{x}\left(1-\cos \theta \right)+u_{z}\sin \theta &\cos \theta +u_{y}^{2}\left(1-\cos \theta \right)&u_{y}u_{z}\left(1-\cos \theta \right)-u_{x}\sin \theta \\u_{z}u_{x}\left(1-\cos \theta \right)-u_{y}\sin \theta &u_{z}u_{y}\left(1-\cos \theta \right)+u_{x}\sin \theta &\cos \theta +u_{z}^{2}\left(1-\cos \theta \right)\end{bmatrix}}
\end{align}
The basic idea to derive this matrix is dividing the problem into a few known simple steps:
\begin{enumerate}
\item First rotate the given axis and the point such that the axis lies in one of the coordinate planes ($xy$, $yz$ or $zx$).
\item Then rotate the given axis and the point such that the axis is aligned with one of the two coordinate axes for that particular coordinate plane ($x$, $y$ or $z$).
\item Use one of the fundamental rotation matrices (about the $x$-, $y$-, or $z$-axis) to rotate the point depending on the coordinate axis with which the rotation axis is aligned.
\item Reverse rotate the axis-point pair such that it attains the final configuration as that was in the second step, in other words, undo the second step.
\item Reverse rotate the axis-point pair which was done in the first step; undo step one. 
\end{enumerate}

Mathematically, the above procedure can be written as: 
\begin{align}
    R=(\cos \theta )\,I+(\sin \theta )\,[\mathbf {u} ]_{\times }+(1-\cos \theta )\,(\mathbf {u} \otimes \mathbf {u} )
\end{align}
where
\begin{align}
    \mathbf {u} \otimes \mathbf {u} =\mathbf {u} \mathbf {u} ^{\mathsf {T}}={\begin{bmatrix}u_{x}^{2}&u_{x}u_{y}&u_{x}u_{z}\\[3pt]u_{x}u_{y}&u_{y}^{2}&u_{y}u_{z}\\[3pt]u_{x}u_{z}&u_{y}u_{z}&u_{z}^{2}\end{bmatrix}},\qquad [\mathbf {u} ]_{\times }={\begin{bmatrix}0&-u_{z}&u_{y}\\[3pt]u_{z}&0&-u_{x}\\[3pt]-u_{y}&u_{x}&0\end{bmatrix}}
\end{align}
In $\mathbb{R}^{3}$ the rotation of a vector $\vec x$ around the axis $\mathbf{u}$ by an angle $\theta$ can be written as:
\begin{align}
    R_{\mathbf{u}}(\theta) \vec x = \mathbf{u} (\mathbf{u} \cdot \vec x) + \cos(\theta)(\mathbf{u} \times \vec x) \times \mathbf{u} + \sin(\theta)(\mathbf{u} \times \vec x)
\end{align}

\subsection{Euler-Rodrigues parameters}
Three-dimensional rotations can be represented by a variety of parametrizations each suitable for applications in different domains \cite{NMR_ERP}. The Euler-Rodrigues parametrization provides a way of performing a rotation as a Cartesian transformation. A central rotation about a rotation axis for a point at $x^{\prime}$ is represented by four real numbers $a,b,c,d$ as 
\begin{equation}
    \vec{x}^{\prime}=\begin{pmatrix} a^{2}+b^{2}-c^{2}-d^{2} & 2(b c-a d) & 2(b d+a c) \\ 2(b c+a d) & a^{2}+c^{2}-b^{2}-d^{2} & 2(c d-a b) \\ 2(b d-a c) & 2(c d+a b) & a^{2}+d^{2}-b^{2}-c^{2}\end{pmatrix} \vec{x}, \label{ER_matrix}
\end{equation}
subject to the constraint $a^{2}+b^{2}+c^{2}+d^{2}=1$. The rotation can be uniquely determined by the rotation angle $\theta_r$ and axis of rotation $(n_x,n_y,n_z)$ by the following equations for $a,b,c$ and $d$.
\begin{equation}
    a  = \cos(\theta_r/2) \hspace{5 mm} 
    b =  n_x \sin(\theta_r/2)  \hspace{5 mm} 
    c =  n_y \sin(\theta_r/2) \hspace{5 mm}
    d = n_z \sin(\theta_r/2)
    \label{ER_sphParam}
\end{equation}

\subsection{Basic quantum operations}

A qubit is the fundamental unit of information in quantum computation. This abstract object can take on states in a Hilbert space spanned by two basis vectors $|0\rangle$ and $|1\rangle$ which correspond to the column vectors $\begin{pmatrix}1 & 0 \end{pmatrix}^\mathsf{T}$ and $\begin{pmatrix}0 & 1\end{pmatrix}^\mathsf{T}$, respectively. Thus, the general state of a qubit is $\alpha|0\rangle+\beta|1\rangle$ with $\alpha, \beta \in \mathbb{C}$ such that $|\alpha|^2+|\beta|^2=1$. When measured, qubits collapse to one of the basis states, which are referred to as the computational basis states. The probabilities of collapsing to $|0\rangle$ and $|1\rangle$ are $|\alpha|^2$ and $|\beta|^2$, respectively. Hence the need for the constraint that these two quantities add up to unity. 

To act on qubits, we apply operators which can be represented as unitary matrices. Some of the most common single-qubit gates are the rotation gates, which can be obtained by taking the exponentials of the Pauli matrices. The Pauli matrices are defined as shown below.
\begin{equation}
    X = \sigma_1 = \begin{pmatrix} 0 & 1 \\ 1 & 0 \end{pmatrix}, \quad Y = \sigma_2 = \begin{pmatrix} 0 & -i \\ i & 0 \end{pmatrix}, \quad Z = \sigma_3 = \begin{pmatrix} 1& 0 \\ 0 & -1\end{pmatrix}
\end{equation}
In some cases, the identity matrix is considered to be the zeroth Pauli matrices, defined as
\begin{equation}
    I = \sigma_0 = \begin{pmatrix}1&0\\0&1\end{pmatrix}.
\end{equation}
Applying these gates to a qubit works as matrix multiplication. Given a general state $|\psi\rangle=\alpha|0\rangle+\beta|1\rangle$, we can apply the $X$ gate to it by doing $X|\psi\rangle$; which is simply matrix-vector multiplication. We can clearly see that $X(\alpha|0\rangle+\beta|1\rangle)=\beta|0\rangle+\alpha|1\rangle$ thanks to the columns of $X$. Note that the $X$ gate flips the amplitudes of the two basis states, so it is commonly referred to as the bit-flip gate. The effect of a gate will not always be this easy to figure out, but it always boils down to matrix-vector multiplication. 

Say you have two unitary matrices $U_0$ and $U_1$. If you want to apply $U_0$ followed by $U_1$ to the state $|\psi\rangle$, you would compute $U_1U_0|\psi\rangle$. Then, you could proceed by doing two matrix-vector multiplications or one matrix-matrix multiplication and one matrix-vector multiplication. Basically, the unitary matrices and statevectors follow the usual rules of linear algebra. 

With the Pauli matrices, we can now define the rotation gates. These represent a rotation of $\theta$ radians about one of the three axes in the Bloch sphere, which helps visualize one qubit quantum states. These gates are the following.
\begin{gather}
    R_x(\theta) = \exp\left(-i X \frac{\theta}{2}\right) = \begin{pmatrix} \cos(\theta/2) & -i\sin(\theta/2) \\ -i\sin(\theta/2) & \cos(\theta/2)
    \end{pmatrix} \label{eq:rot_x} \\
    R_y(\theta) = \exp\left(-i Y \frac{\theta}{2}\right) = \begin{pmatrix} \cos(\theta/2) & -\sin(\theta/2) \\ \sin(\theta/2) & \cos(\theta/2)
    \end{pmatrix} \label{eq:rot_y} \\
    R_z(\theta) = \exp\left(-i Z \frac{\theta}{2}\right) = \begin{pmatrix} e^{-i \theta / 2} & 0 \\ 0 & e^{i \theta / 2} \end{pmatrix} \label{eq:rot_z}
\end{gather}
These gates are fundamental to quantum computation since they are the easily implementable in current hardware. An important property about these gates is that their inverse is defined by a rotation about the same axis but with a negative angle. This is to say that $R^{-1}_i(\theta)=R^{\dagger}_i(\theta)=R_i(-\theta)$. Also, having two or more consecutive rotations about the same axis is equivalent to a single rotation about that axis with the angle being the total sum of the angles of the individual rotations. It should be noted that the Pauli matrices follow the equality:
\begin{equation}
    X^2=Y^2=Z^2=-iXYZ=I \label{eq:pauli_identities}
\end{equation}
Thanks to the fact that each Pauli matrix squared is equal to the identity, we can expand the exponential in the equations for the rotation operators as $\exp(-i \sigma_i \nicefrac{\theta}{2})=\cos(\nicefrac{\theta}{2})I - i\sin(\nicefrac{\theta}{2})\sigma_i$. Equation \eqref{eq:pauli_identities} also gives us a blueprint to calculate the product of any pair of Pauli matrices. To do this take the last identity ($-iXYZ=I$) and do the necessary left and right multiplications until the right hand side of the equation is the pair (or triplet) you want to compute. Additionally, a general single-qubit unitary $U$ can be expressed as $e^{i \delta}R_z(\alpha)R_y(\theta)R_z(\beta)$ \cite{Barenco_1995}, where the first term represents a global phase that becomes relevant when working with multi-qubit systems.

\section{Rotation around an arbitrary axis}

\subsection{Rotation scheme}\label{sec:Bloch_sph}

A general single-qubit state can be defined as $|\psi\rangle = e^{i \delta}\left( e^{-i\phi/2}\cos(\theta/2)|0\rangle+e^{i\phi/2}\sin(\theta/2)|1\rangle \right)$, where $e^{i \delta}$ is an undetectable global phase and  we can interpret the other two angles as spherical coordinates of the form $(1, \theta, \phi)$ \cite{Shende_2006}. Which in turn represents the point in the surface of the Bloch sphere corresponding to the state $|\psi\rangle$. Thinking about this state in geometrical terms, we know that it is $\theta$ radians away from the unit vector in the positive $z$-axis direction and $\phi$ radians away from the $x$-axis unit vector. 

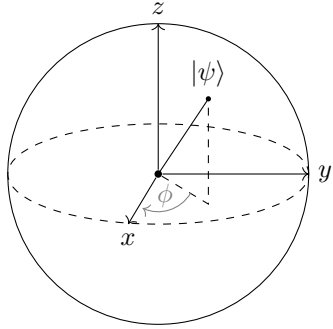
\begin{figure}[H]
    \centering
    \begin{tikzpicture}
        \def\r{2}
    
        \draw (0,0) node[circle,fill,inner sep=1] (orig) {} -- (\r/3,\r/2) node[circle,fill,inner sep=0.7,label=above:$|\psi\rangle$] (a) {};
        \draw[dashed] (orig) -- (\r/3,-\r/5) node (phi) {} -- (a);
    
        \draw (orig) circle (\r);
        \draw[dashed] (orig) ellipse (\r{} and \r/3);
    
        \draw[->] (orig) -- ++(-\r/5,-\r/3) node[below] (x1) {$x$};
        \draw[->] (orig) -- ++(\r,0) node[right] (x2) {$y$};
        \draw[->] (orig) -- ++(0,\r) node[above] (x3) {$z$};
    
        \pic [draw=gray,text=gray,<-,"$\phi$"] {angle = x1--orig--phi};
    \end{tikzpicture}
    \caption{Rotating $-\phi$ radians around the $z$-axis}
    \label{fig:r_z_rotation}
\end{figure}

Therefore, if we want to transform the $|\psi\rangle$ state into $|0\rangle$, which has spherical coordinates $(1, 0, 0)$, we can do so by applying $R_z(-\phi)$ followed by $R_y(-\theta)$. This procedure can be visually represented with rotations in the Bloch sphere. First, as shown in Figure \ref{fig:r_z_rotation}, a rotation around the $z$-axis places the statevector $|\psi\rangle$ in the $xz$ plane. 

After this rotation, the new state is $|\psi^\prime\rangle = e^{i\delta}\left( \cos(\theta/2)|0\rangle + \sin(\theta/2)|1\rangle\right)$ which has spherical coordinates $(1, \theta, 0)$. Then, the $R_y(-\theta)$ gate rotates $|\psi^\prime\rangle$ about the $y$-axis to finally place it in the north pole of the Bloch sphere.  

\begin{figure}[H]
    \centering
    \begin{tikzpicture}
        \def\r{2}
    
        \draw (0,0) node[circle,fill,inner sep=1] (orig) {} -- (\r/2,\r*0.866) node[circle,fill,inner sep=0.7,label=above:$|\psi\rangle$] (a) {};
        \draw[dashed] (orig) -- (\r/2,0) node (phi) {} -- (a);
    
        \draw (orig) circle (\r);
        \draw[dashed] (orig) ellipse (\r{} and \r/3);
    
        \draw[->] (orig) -- ++(-\r/5,-\r/3) node[below] (x1) {$y$};
        \draw[->] (orig) -- ++(\r,0) node[right] (x2) {$x$};
        \draw[->] (orig) -- ++(0,\r) node[above] (x3) {$z$};
    
        \pic [draw=gray,text=gray,->,"$\theta$"] {angle = a--orig--x3};
    \end{tikzpicture}
    \caption{Rotating $-\theta$ radians around the $y$-axis ($x$ and $y$ axis exchanged for clarity)}
    \label{fig:r_y_rotation}
\end{figure}
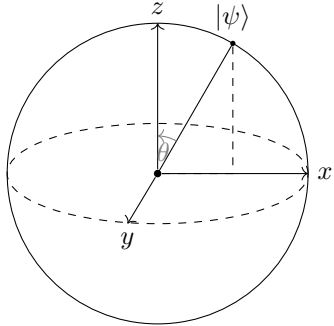


Thus, the final state after these two rotations is $|\psi^{\prime\prime}\rangle = e^{i \delta}|0\rangle$, which is the usual ground state up to a negligible global phase. In mathematical terms, the procedure goes as follows.
\begin{align}
\begin{split}
    R_y(-\theta)&R_z(-\phi)|\psi\rangle \\
    &= R_y(-\theta)R_z(-\phi) \left[e^{i \delta}\left( e^{-i\phi/2}\cos\left(\nicefrac{\theta}{2}\right)|0\rangle+e^{i\phi/2}\sin\left(\nicefrac{\theta}{2}\right)|1\rangle \right)\right] \\
    &= e^{i \delta} R_y(-\theta)  \left( e^{i\phi/2}e^{-i\phi/2}\cos\left(\nicefrac{\theta}{2}\right)|0\rangle+e^{-i\phi/2}e^{i\phi/2}\sin\left(\nicefrac{\theta}{2}\right)|1\rangle \right)  \\
    &=e^{i \delta}R_y(-\theta) \left(\cos\left(\nicefrac{\theta}{2}\right)|0\rangle+\sin\left(\nicefrac{\theta}{2}\right)|1\rangle \right) \\
    &= e^{i \delta} \left[ \cos\left(\nicefrac{\theta}{2}\right)\left(\cos\left(\nicefrac{-\theta}{2}\right)|0\rangle + \sin\left(\nicefrac{-\theta}{2}\right)|1\rangle\right) + \sin\left(\nicefrac{\theta}{2}\right)\left(\cos\left(\nicefrac{-\theta}{2}\right)|1\rangle-\sin\left(\nicefrac{-\theta}{2}\right)|0\rangle\right) \right] \\
    &= e^{i \delta}\left[ \cos(\nicefrac{\theta}{2})\cos\left(\nicefrac{-\theta}{2}\right)|0\rangle - \sin\left( \nicefrac{\theta}{2} \right)\sin\left(\nicefrac{-\theta}{2}\right)|0\rangle\right] \\
    &= e^{i\delta}\cdot\nicefrac{1}{2}\cdot\left[ (\cos(\theta)+1)|0\rangle-(\cos(\theta)-1)|0\rangle \right] \\
    &= e^{i\delta}|0\rangle
\end{split}
\end{align}
This allows us to define a way of rotating an arbitrary quantum state around an arbitrary axis $\vec n = (n_x, n_y, n_z)$ with rotating angle $\theta_r$, thinking in terms of the Bloch sphere. First, of all, we are going to assume the vector $\vec n$ is normalized and thus we can write it in spherical coordinates as $(1, \theta_n, \phi_n)$. Given this, we can define a scheme for rotation around $\vec n$ as follows:
\begin{enumerate}
    \item  Perform the necessary rotations so $(1, \theta_n, \phi_n)$ is now at $(1, 0, 0)$. This is done by applying $R_z(-\phi_n)$ followed by $R_y(-\theta_n)$.
    \item Now that $\vec n$ is at $|0\rangle$, we can perform our desired rotation by doing $R_z(\theta_r)$. This works since the original rotation axis is now aligned with the $z$-axis.
    \item Perform rotations to get everything to its original spot, in other words, do the inverse of step 1. This is done by applying $R_y(\theta_n)$ followed by $R_z(\phi_n)$.
\end{enumerate}
In geometric terms, this can be interpreted as translating the rotation axis into the $z$-axis, performing the desired rotation about the $z$-axis, and finally bringing the original rotation axis back to its original spot. The complete operation for this procedure is $R_z(\phi_n)R_y(\theta_n)R_z(\theta_r)R_y(-\theta_n)R_z(-\phi_n)$.

\subsection{Euler-Rodrigues parameters from circuit unitary}

The unitary operation corresponding to the scheme described above can also be written as $R_{\vec n}(\theta_r) = \exp(-i \cdot \nicefrac{\theta_r}{2} \cdot \vec n \cdot \vec \sigma)$, where $\sigma$ is a vector of Pauli matrices. Expanding the exponent gives us the following expression for the unitary:
\begin{align}
\begin{split}
    R_{\vec n}(\theta_r) = \exp\left(-i \cdot \frac{\theta_r}{2} \cdot \vec n \cdot \vec \sigma\right) &= \cos\left(\frac{\theta_r}{2}\right)I - i\sin\left(\frac{\theta_r}{2}\right)\vec n \cdot \vec \sigma \\
    &= \cos\left(\frac{\theta_r}{2}\right)I - i\sin\left(\frac{\theta_r}{2}\right)(n_x \sigma_1 + n_y \sigma_2 + n_z \sigma_3) \label{eq:rot_n_expanded}
\end{split}
\end{align}
We can show that equation \eqref{eq:rot_n_expanded} is equal to the five gate sequence $R_z(\phi_n)R_y(\theta_n)R_z(\theta_r)R_y(-\theta_n)R_z(-\phi_n)$ by using the identities introduced in equations \eqref{eq:rot_x}, \eqref{eq:rot_y} and \eqref{eq:rot_z}. We start by expanding the expression for the unitary in the middle, which gives us:
\begin{align}
    R_z(\phi_n)R_y(\theta_n)\left(\cos\left(\frac{\theta_r}{2}\right)I - \sin\left(\frac{\theta_r}{2}\right)Z\right)R_y(-\theta_n)R_z(-\phi_n) \label{eq:mid_rz_expanded}
\end{align}
Now, we are going to briefly show that $R_y(\theta)ZR_y(-\theta)=\cos(\theta)Z+\sin(\theta)X$ by using Pauli products that can be derived from equation \eqref{eq:pauli_identities}. 
\begin{align}
\begin{split}
    R_y(\theta)ZR_y(-\theta) &= \left(\cos\left(\frac{\theta}{2}\right)I-i\sin\left(\frac{\theta}{2}\right)Y\right) Z \left(\cos\left(\frac{\theta}{2}\right)I+i\sin\left(\frac{\theta}{2}\right)Y\right) \\
    &= \cos^2\left(\frac{\theta}{2}\right)Z + i\cos\left(\frac{\theta}{2}\right)\sin\left(\frac{\theta}{2}\right)ZY - i\cos\left(\frac{\theta}{2}\right)\sin\left(\frac{\theta}{2}\right)YZ + \sin^2\left(\frac{\theta}{2}\right)YZY \\
    &= \left(\cos^2\left(\frac{\theta}{2}\right)-\sin^2\left(\frac{\theta}{2}\right)\right)Z + 2\cos\left(\frac{\theta}{2}\right)\sin\left(\frac{\theta}{2}\right)X \\
    &= \cos(\theta)Z + \sin(\theta)X
\end{split}
\end{align}
The first step was to expand the rotation matrices using the generalized Euler formula. Then, we carried matrix multiplication to get rid of the parentheses. After that, we used the Pauli pair identities to factorize the expression and finally we were able to use regular trigonometric identities to get the result we wanted. Using the same outline, we can prove that $R_z(\theta)ZR_z(-\theta)=Z$ and $R_z(\theta)XR_z(-\theta)=\cos(\theta)X+\sin(\theta)Y$. And remember that $R_i(\theta)R_i(\phi)=R_i(\theta+\phi)$. We can now continue with our original goal. Using the identities stated above, we can continue from equation \eqref{eq:mid_rz_expanded} as follows:
\begin{align}
\begin{split}
    R_z(\phi_n)R_y(\theta_n)&\left(\cos\left(\frac{\theta_r}{2}\right)I - \sin\left(\frac{\theta_r}{2}\right)Z\right)R_y(-\theta_n)R_z(-\phi_n) \\ &= R_z(\phi_n)\left[ \cos\left(\frac{\theta_r}{2}\right)I - i \sin\left(\frac{\theta_r}{2}\right)\left(\cos(\theta_n)Z+\sin(\theta_n)X\right) \right]R_z(-\phi_n) \\
    &= \cos\left(\frac{\theta_r}{2}\right)I - i\sin\left(\frac{\theta_r}{2}\right)\left[\cos(\theta_n)Z + \sin(\theta_n)(\cos(\phi_n)X+\sin(\phi_n)Y)\right] \\
    &= \cos\left(\frac{\theta_r}{2}\right)I - i\sin\left(\frac{\theta_r}{2}\right)\left(\sin(\theta_n)\cos(\phi_n)X + \sin(\theta_n)\sin(\phi_n)Y + \cos(\theta_n)Z\right)
\end{split}
\end{align}
This expression already looks very similar to \eqref{eq:rot_n_expanded}, and in fact is equivalent. The coefficients in front of the Pauli matrices are equal to the equations for converting spherical coordinates to Cartesian coordinates, with the radius $r$ set to $1$ (which is our case thanks to the normalization of $\vec n$). Thus, we have proved that the five-qubit sequence we use to rotate in the Bloch sphere is equal to the rotation operator obtained by exponentiation.

From the background, recall that a rotation of $\theta$ radians about an axis $\vec n$ can be defined in terms of the Euler-Rodrigues parameters $a, b, c$ and $d$ as defined in \cite{NMR_ERP}. We can extract these from the unitary that corresponds to our circuit. To do this, first remember that the Pauli matrices (including the identity) form a basis for the $2 \times 2$ Hermitian matrices. So, any unitary operation can be written as follows:
\begin{equation}
    U = \alpha\sigma_1 + \beta\sigma_2 + \gamma\sigma_3 + \delta I
\end{equation}
Remember that the trace of a square matrix is defined as the sum of elements in its main diagonal. Thus, we have the following properties:
\begin{gather}
    \mathrm{Tr}(U\sigma_{1})/2 = \alpha\hspace{5mm} \mathrm{Tr}(U\sigma_{2})/2 = \beta\hspace{5mm} \mathrm{Tr}(U\sigma_{3})/2 = \gamma\hspace{5mm} \mathrm{Tr}(U)/2 = \delta \label{eq:trace_properties}
\end{gather}
From expanding the parenthesis in \eqref{eq:rot_n_expanded}, we can easily see that $\alpha, \beta, \gamma$ and $\delta$ for the rotation we defined correspond directly to the parameters $a, b, c$ and $d$ that we are looking to extract. Applying the properties in \eqref{eq:trace_properties} to the unitary of $R_{\vec n}(\theta_r)$, we get the following relations:
\begin{align}
\begin{split}
    \delta = \cos(\theta_r/2) = a \hspace{5 mm} 
    \alpha = -i n_x \sin(\theta_r/2) = -ib \\
    \beta = -i n_y \sin(\theta_r/2) = -ic \hspace{5 mm}
    \gamma = -i n_z \sin(\theta_r/2) = -id 
\end{split}
\end{align}
Therefore, the information that describes a rotation using the Euler-Rodrigues formula can be extracted from the proposed circuit with minimal post-processing. However, for this to be completely accurate we rely on simulators or assume a noise-free quantum device.





\subsection{Tomography for parameter extraction}\label{sec:tomo_param_extraction}

The procedure described on the previous subsection to extract the rotation parameters assumes complete access to the circuit unitary, which is unrealistic when running experiments on a real device. When working on real devices, we can perform process tomography \cite{Chuang_1997} to characterize our circuit's process using the Choi-matrix representation. Then, we obtain the Kraus operators through the eigenvalue decomposition of the Choi-matrix $C = U S V^{\dagger}$, where $U$ and $V$ are unitary matrices and $S$ is a diagonal matrix. For the one qubit case, which we consider here, $C$ is a $4\times4$ matrix. If the process is completely-positive, as in this case, the left and right Kraus operators are identical. So, we compute the Kraus operators as
\begin{equation}
    K_i = \sqrt{s_{ii}}\begin{bmatrix}u_{1i} & u_{3i} \\ u_{2i} & u_{4i}  \end{bmatrix}.
\end{equation}
If all the values of $S$ are zero except for one, then the only non-zero Kraus operator satisfies $K_i^{\dagger} K_i= K_i K_i^{\dagger} = I$ and therefore is unitary. In this case, that single operator is also the unitary operator we are looking to reconstruct. To obtain the data needed for this reconstruction we create and run $4^n3^n$ circuits a large amount of times each. That amount of circuits is needed because they consist of preparing each of the four minimal Pauli basis eigenstates ($|0\rangle, |1\rangle, |+\rangle, |+i\rangle$), sending them through the circuit and measuring in each of the three Pauli basis. 

\begin{figure}
    \centering
    \includegraphics[width=.7\linewidth]{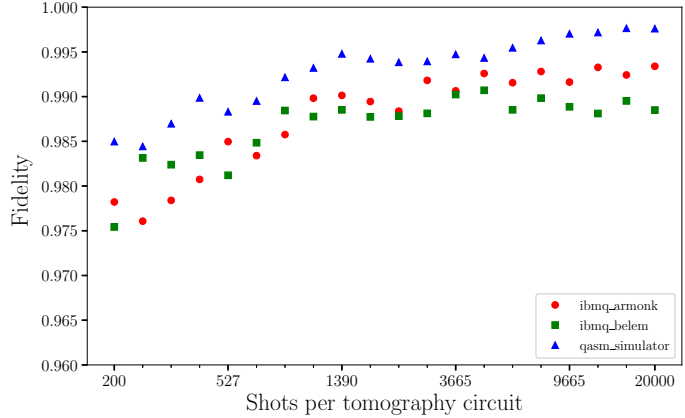}
    \caption{Fidelity of reconstructed unitary against shots used in process tomography}
    \label{fig:process_tomo_fidelity}
\end{figure}

For testing purposes, we generated twenty random rotations and ran process tomography on the corresponding rotation circuits with twenty logarithmically spaced numbers of shots between two hundred and twenty thousand. This was run in a simulator and two real quantum devices. After reconstructing the unitary through tomography, we calculate the gate fidelity between that and the exact unitary and plot the results in Figure \ref{fig:process_tomo_fidelity}.

\section{Rotating a vector}

\subsection{Initializing and rotating arbitrary vectors}

Using the procedure described on the last section combined with an encoding procedure, we can rotate any normalized vector. Suppose we have an input vector with spherical coordinates $(1, \theta_i, \phi_i)$ and we want to rotate $\theta_r$ radians about $\vec n$. Doing this is pretty simple: first we need to initialize a single qubit to the quantum state corresponding to $(1, \theta_i, \phi_i)$ which can be done by applying two rotation gates, and then we apply the rotation scheme described before. What we are left with is the quantum state with spherical coordinates $(1, \theta_f, \phi_f)$, up to a global phase.

In the last section, we looked into the unitary matrix corresponding to the quantum circuit in order to extract useful information about the Euler-Rodrigues parameters. In this case, we want to gain information about the rotated vector and not the rotation matrix.

\subsection{Extracting rotated vector}
\begin{figure}
\centering
\begin{subfigure}{.5\textwidth}
  \centering
  \includegraphics[width=.65\linewidth]{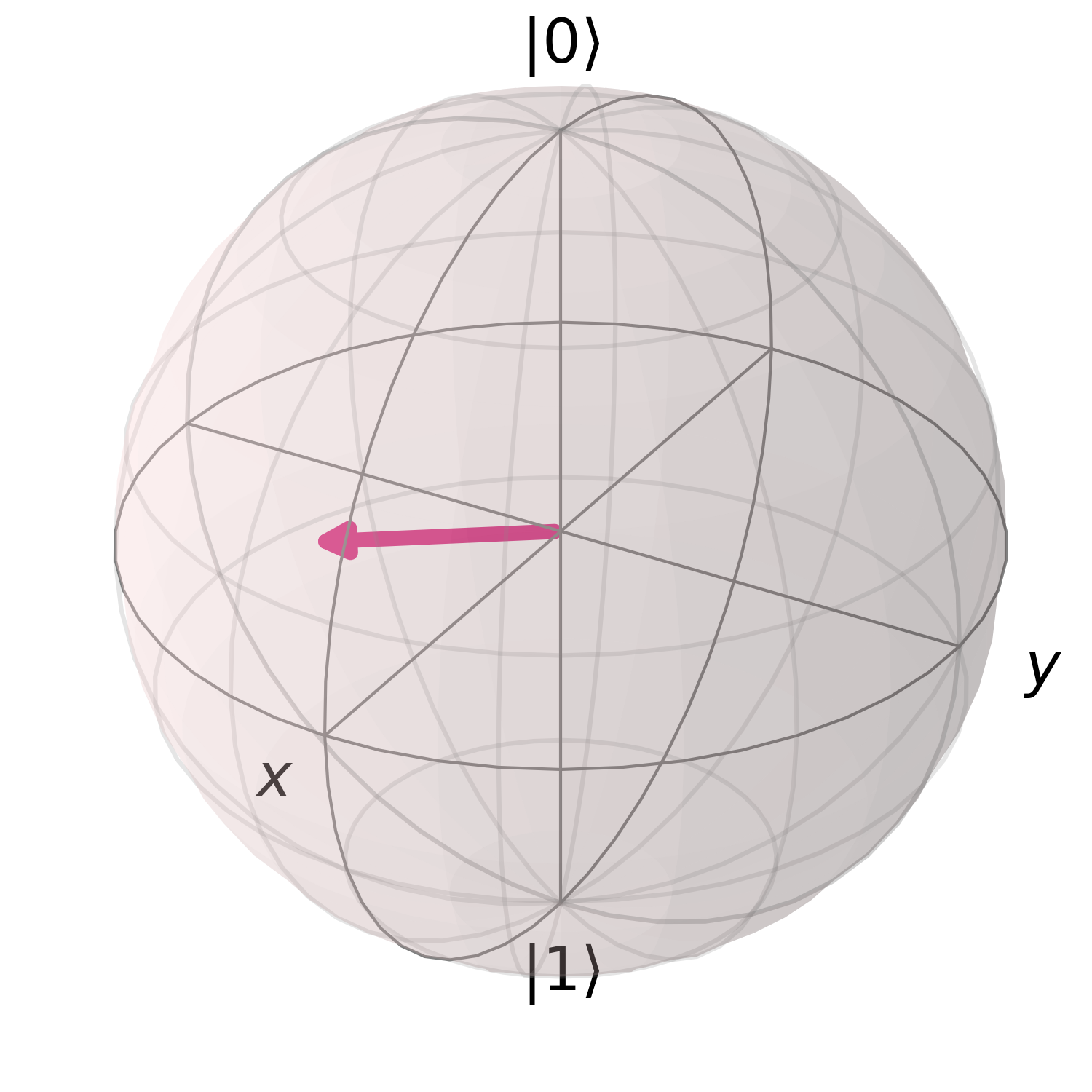}
  \caption{Classical calculation of the Euler-Rodrigues matrix.}
  \label{fig:rot_classical}
\end{subfigure}%
\begin{subfigure}{.5\textwidth}
  \centering
  \includegraphics[width=.65\linewidth]{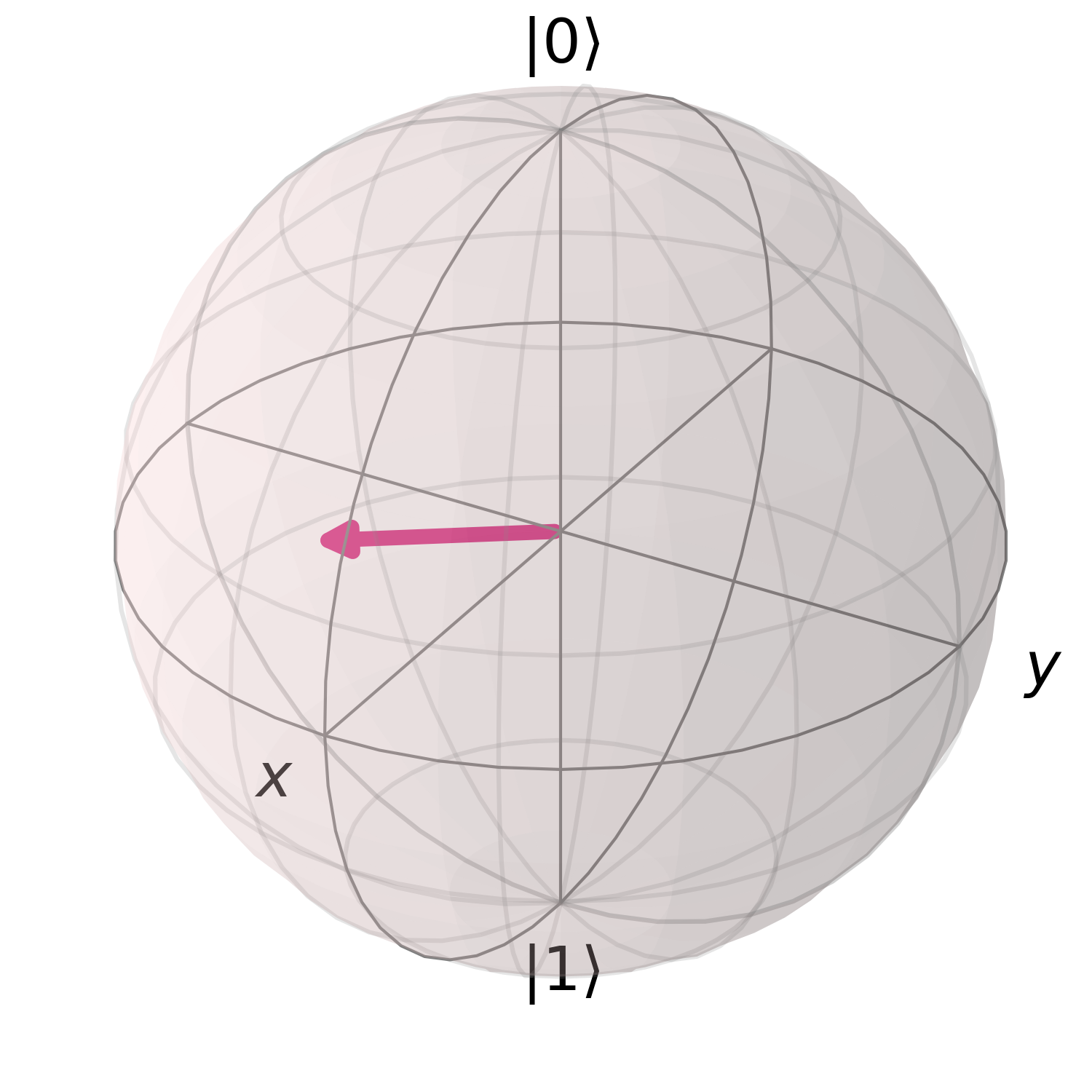}
  \caption{Quantum rotation circuit and tomography simulation.}
  \label{fig:rot_quantum}
\end{subfigure}
\caption{Result of a $\pi/4$ radians rotation of $[1, 0, 1]$ about $[1, 0.5, 0.5]$ (both vectors normalized to unit lenght). Figures generated with Qiskit \cite{Qiskit_2021}.}
\label{fig:bloch_rotations}
\end{figure}

To extract the rotated vector, we use a version of the quantum state tomography procedure introduced in \cite{Smolin_2012}. The procedure given in this paper is quite simple and for a single qubit needs only three measurement basis: $X$, $Y$, and $Z$. To obtain high fidelity results, we run the circuit a large amount of times for each of these three measurement basis. The measurement results allow us to apply the algorithm introduced by Smolin to get the maximum likelihood density matrix. We then compute the eigenvalues $\lambda_i$ and eigenvectors $|\lambda_i\rangle$ of this density matrix so we can get the statevector $|\psi\rangle = \sum_i\lambda_i|\lambda_i\rangle$. Finally, we can plot this state in the Bloch sphere and compare it to the vector we obtain rotating classically to confirm they are equivalent as showed in Figure \ref{fig:bloch_rotations}. Visually, this is a really good indicator that the vector obtained through tomography is the desired rotated vector.

\begin{figure}
    \centering
    \includegraphics[width=.7\linewidth]{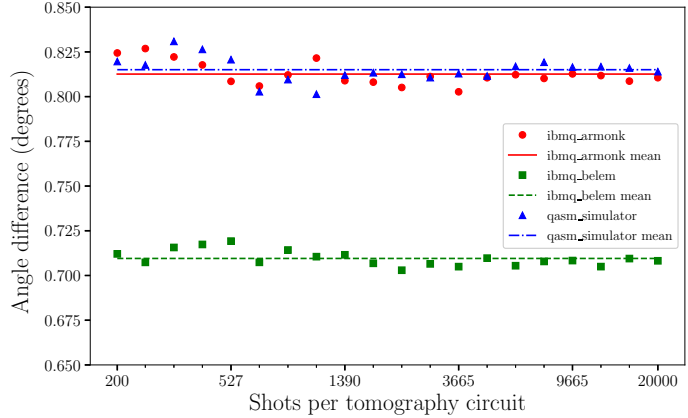}
    \caption{Angle difference between classical and quantum rotation against shots used in tomography}
    \label{fig:angle_diff}
\end{figure}

A more precise metric we can use to confirm the accuracy of our method is the angle between the classical and quantum rotated vector, which becomes smaller as the number of shots in the tomography is increased. After converting both vectors into rectangular coordinates, we can use the well-known formula for the angle between two vectors: $\cos(\theta)=(\vec{x} \cdot \vec{y}) / (|\vec x||\vec y|)$. We use the same randomly generated rotations and devices as in Section \ref{sec:tomo_param_extraction} to test state tomography on our rotation scheme, using the angle between the extracted vector and the classically calculated vector as a metric. These results are showed in Figure \ref{fig:angle_diff}.
\section{Multi-qubit rotations}\label{sec:multi_qubit_rotations}

\subsection{Encoding initial vectors}

To apply the same rotation to multiple vectors, we resort to using $n+1$ qubits, which will allow us to rotate up to $2^n$ vectors at once. We use $n$ qubits as control qubits and the remaining qubit will act as a data qubit in which we will store all the vectors before and after rotating them. To illustrate our procedure, we can describe the case where $n=2$. 

First, we are going to put the $n$ control qubits in an equal superposition by applying a Hadamard to each qubit. This is equivalent to applying $H^{\otimes n}$ and for $n=2$ gives us the state below.
\begin{equation}
    (H^{\otimes 2}|00\rangle)|0\rangle = (|00\rangle+|01\rangle+|10\rangle+|11\rangle)|0\rangle
\end{equation}
As you can see, the data qubit is left undisturbed in this first step and we have an equal superposition of the control qubits. 

Then, we take the $n$-controlled versions of the $R_y$ and $R_z$ gates. As explained in the single-qubit case, these rotation gates are used to encode a vector and then rotate it. The controlled versions are going to help us encode a vector for each basis state of the $n$ control qubits. Since we have one vector for each basis state of a $n$-qubit register, we are able to encode $2^n$ vectors. 

To go ahead with the encoding, first we calculate the rotation angles $\{\theta_i, \phi_i\}$ for each $i$-th vector where $1 \leq i \leq 2^n$ just as described in the single-qubit case. Then, we apply the controlled gates with the control qubits surrounded by $X$ gates depending on the value of $i$ and with the data qubit as target. When $i = 1$, no $X$ gates are used. When $i = 2$, an $X$ gate in the first qubit is used. And the pattern continues: the $X$ gates are used to encode the binary representation of $i-1$. A circuit for our example would look as follows:

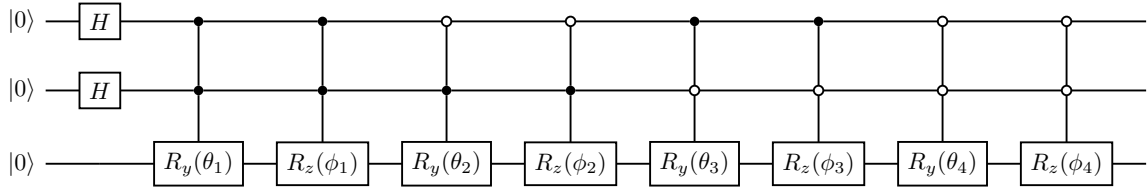
\begin{figure}[H]
    \centering
    \begin{tikzpicture}
        \node[scale=0.9]{
        \begin{quantikz}
            \lstick{$|0\rangle$} & \gate{H} & \ctrl{1} & \ctrl{1} & \octrl{1} & \octrl{1} & \ctrl{1} & \ctrl{1} & \octrl{1} & \octrl{1} & \qw \\
            \lstick{$|0\rangle$} & \gate{H} & \ctrl{1} & \ctrl{1} & \ctrl{1} & \ctrl{1} & \octrl{1} & \octrl{1} & \octrl{1} & \octrl{1} & \qw \\
            \lstick{$|0\rangle$} & \qw & \gate{R_y(\theta_1)} & \gate{R_z(\phi_1)} & \gate{R_y(\theta_2)} & \gate{R_z(\phi_2)} & \gate{R_y(\theta_3)} & \gate{R_z(\phi_3)} & \gate{R_y(\theta_4)} & \gate{R_z(\phi_4)} & \qw
    \end{quantikz}};
    \end{tikzpicture}
    \caption{Encoding $4$ vectors using $2$ control qubits}
    \label{fig:multi_encoding_example}
\end{figure}

In this figure, white controls are normal controls surrounded by $X$ gates, i.e., they are triggered by $|0\rangle$ instead of $|1\rangle$. The state of the circuit above evolves as
\begin{align}
\begin{split}
    |00\rangle |0\rangle &\xrightarrow[]{H^{\otimes 2}} (|00\rangle + |01\rangle +|10\rangle +|11\rangle)|0\rangle = |00\rangle|0\rangle + |01\rangle|0\rangle + |10\rangle|0\rangle + |11\rangle|0\rangle \\ &\xrightarrow[]{C_{11}R_z C_{11}R_y} |00\rangle |0\rangle + |01\rangle|0\rangle + |10\rangle|0\rangle + |11\rangle|\vec v_4\rangle \\ &\xrightarrow[]{C_{10}R_z C_{10}R_y} |00\rangle |0\rangle + |01\rangle|0\rangle + |10\rangle|\vec v_3\rangle +|11\rangle|\vec v_4\rangle \\ &\xrightarrow[]{C_{01}R_z C_{01}R_y} |00\rangle |0\rangle + |01\rangle|\vec v_2\rangle + |10\rangle|\vec v_3\rangle + |11\rangle|\vec v_4\rangle \\ &\xrightarrow[]{C_{11}R_z C_{11}R_y} |00\rangle |\vec v_1\rangle + |01\rangle|\vec v_2\rangle + |10\rangle|\vec v_3\rangle +|11\rangle|\vec v_4\rangle \label{eq:final_encoding}
\end{split}
\end{align}
where $|\vec v_i\rangle$ is the $i$-th vector encoded as $R_z(\phi_i)R_y(\theta_i)|0\rangle$.  For general $n$, the encoding procedure would look as follows.
\begin{align}
    |0\rangle^{\otimes n}|0\rangle &\xrightarrow[]{H^{\otimes n}} \sum_{i=1}^{2^n}|i-1\rangle |0\rangle \\ &\xrightarrow[]{C_iR_zC_iR_y} \sum_{i=1}^{2^n}|i-1\rangle |\vec v _i\rangle
\end{align}

\subsection{Rotating the superposition}

Now, we can apply the usual single-qubit rotation scheme to the data qubit such that it rotates all of the qubits we encoded. Remember that this scheme corresponds to applying the gate sequence $R_z(\phi_n)R_y(\theta_n)R_z(\theta_r)R_y(-\theta_n)R_z(-\phi_n)$, where we are rotating $\theta_r$ radians about an axis with spherical coordinates $(1, \theta_n, \phi_n)$. For simplicity, we can compact these gates into a single operator which we will call $R(\theta_n, \phi_n, \theta_r)$. Applying this operator to the state we left off with in equation \eqref{eq:final_encoding}, we get the state
\begin{align}
\begin{split}
    (I\otimes I \otimes R)(|00\rangle |\vec v_1\rangle + |01\rangle|\vec v_2\rangle + |10\rangle|\vec v_3\rangle +|11\rangle|\vec v_4\rangle) &= |00\rangle R|\vec v_1\rangle + |01\rangle R|\vec v_2\rangle + |10\rangle R|\vec v_3\rangle +|11\rangle R|\vec v_4\rangle \\
    &= |00\rangle |\vec r_1\rangle + |01\rangle|\vec r_2\rangle + |10\rangle|\vec r_3\rangle +|11\rangle|\vec r_4\rangle \label{eq:rotated_vectors}
\end{split}
\end{align}
where we use $\vec r_i$ to denote the rotated vectors. For general $n$, the transformation above would look as follows.
\begin{align}
    (I^{\otimes n} \otimes R)\sum_{i=1}^{2^n}|i-1\rangle|\vec v_i\rangle = \sum_{i=1}^{2^n}|i-1\rangle|\vec r_i\rangle
\end{align}
If we want to rotate the encoded vectors by different angles and/or about different vectors, we would need to divide $R$ into two parts: $R_t(\theta_n, \phi_n)$ which corresponds to the initial and final translation (hence the subindex $t$), and the regular rotation about the $z$-axis $R_z(\theta_r)$. This way, we could control these operations on the basis states of the control qubits to perform different rotations to the initial vectors. 

Either way, if we perform the same rotation to all vectors or have any number of different rotations, at the end we end up with a state of the form showed in equation \eqref{eq:rotated_vectors}. The general structure of the data we encoded first doesn't change, but the encoded data is what gets modified. This is convenient as it allows us to extract a desired rotated vector following a very similar procedure to the single-qubit case. 

\subsection{Post-selecting control qubits}

To extract a specific vector, we need to collapse the control qubits to the state corresponding to its index. Remember that the $i$-th vector corresponds to the binary representation of $i-1$ in the control qubits. If we  measure the control qubits as they stand in equation \eqref{eq:rotated_vectors} and post-select, we would only get our desired basis state with $1 / 2^n$ probability, which is clearly not ideal. 

To increase the probability of measuring the state we want and therefore carry on a successful post-selection, we could apply a standard amplitude amplification procedure. More specifically, we would like to mark those states that have our desired index in the control qubits. This could be easily done with a phase oracle consisting of only $Z$ gates. Although the oracle is inexpensive, we would need to apply the initialization procedure with each iteration of the oracle, i.e., $\left\lfloor \frac{\pi}{4} 2^{n/2 }\right \rfloor$ times \cite{Brassard_2002} with $n$ being the number of control qubits. Moreover, the initialization circuit uses $2^{n+1}$ $n$-controlled rotation gates, which are expensive to decompose. 

Sequences of consequent controlled rotation gates with different rotation axis, just as in the circuit in figure \ref{fig:multi_encoding_example}, require $O(2^n)$ nearest-neighbor CNOTs when decomposed \cite{Mottonen_2005}. If all the controlled rotations about the $y-$axis are done before the ones about the $z$-axis (or vice-versa), there is a linear improvement in this decomposition, but the leading term remains exponential. Therefore, an amplitude amplification procedure to increase the success probability of post-selection is expensive with large $n$.

Knowing this, a first instinct would be trying to perform amplitude amplification only on the control qubits, so that the initialization operation would simply be a layer of Hadamard gates. However, entanglement between the control qubits and data qubit forbids such procedure. To see this, suppose Alice has the control qubits and Bob the data qubit and these two parties are physically apart from each other. If amplitude amplification acting only on the control qubits to achieve our goal was possible, Alice could use this to force the data qubit into a state that conveys a message to Bob, allowing faster than light communication. 

Thus, we are left with either an expensive amplitude amplification algorithm or having to deal with low success probability. A possible alternative to this problem is to store the rotated vectors in a QRAM structure \cite{Giovannetti_2008} that would allow easier access to the states stored, at the expense of having to use more spacial resources. Because of the resource constraint, this option may be more well suited for the post-NISQ era. Regardless of the post-selection strategy chosen to collapse the control qubits into the desired index, state or process tomography has to be applied to the state of the data qubit as described in earlier sections of this paper.

\subsection{Application}

This multi-vector rotation scheme could be used to perform rotations of three-dimensional rigid bodies. To do this, follow the encoding and rotation circuit introduced earlier this section using $n = \lceil\log_2 p\rceil$ where $p$ is the number of points in the body. The resulting state would contain the rotated points in the data qubit, which would need to be extracted. Due to the probabilistic nature of quantum circuits and measurement techniques, the resulting points would be an approximation of the exact rotation. The accuracy of this approximation depends on the number of shots used for the tomography.

Since we want to extract all rotated vectors in order to reconstruct the rotated body and not just some specific vectors, the considerations discussed previously about post-selection are not as important in this case. Say we decide to use $k$ shots for the tomography of each vector in order to achieve some arbitrary accuracy. In total, we will need $k \cdot p$ shots of the circuit. This quantity can be upper bounded by $k \cdot 2^n$, which happens when $p$ is a power of $2$. 

Instead of needing to run an amplitude amplification procedure in order to increase the probability of a successful post-selection, we can simply run the circuit $k \cdot 2^n$ times, collapsing the control qubits without any further manipulation. Since the control qubits are in a uniform superposition, i.e., we measure each with probability $1/2^n$, we will get each state approximately $(k \cdot 2^n) \cdot (1/2^n) = k$ times, just as we need. Although we don't get exactly $k$ samples of each state, small deviations in a large number of shots $k$ do not affect the fidelity too much, as shown in figures \ref{fig:angle_diff} and \ref{fig:process_tomo_fidelity}.

It is a good time to discuss a small nuance when we talk about shots in a tomography procedure. Throughout this paper, the number of shots stated for some tomography, refer to the shots in a single circuit of said tomography. This is done since process and state tomography require different number of circuits: $3^n4^n$ and $3^n$ where $n$ is the number of qubits, respectively. So, for the whole tomography, the total number of shots is obtained by multiplying by these factors. Since we only perform tomography on single-qubit states, this only constitutes a linear increase in the total number of shots, therefore the asymptotic behavior remains the same. 

Another important point to address is that rotating and extracting multiple points using this multi-qubit scheme does not present an advantage over using the single-qubit procedure for each point. However, it may present itself useful when we do not wish to extract each single point, but learn something about their structure without having to know the exact state of each point. 

\section{Application Example of Single Qubit Rotation}
In this section, we set out to demonstrate an example of the single qubit rotation scheme as described above. 
\subsection{Vector Considerations}
Let the axis of rotation be represented by the line passing through origin and another point $\vec{b}$ as $\vec{r}=\vec{0}+\mathrm{t} .(\vec{b}-\vec{0}) = t.\vec{b}$. In this example we consider $\vec{b} = (2 , 1, 1)$. Hence the axis when represented by a unit vector will be given as $\vec{n} = \hat{b} = \frac{1}{\sqrt{6}}(2, 1, 1)$
Let the unit vector to be rotated be chosen arbitrarily as $\vec{x} = \frac{1}{\sqrt{3}}(1 , 1, 1)$ and the angle of rotation be considered as $90^{\circ}$. Our goal is to find the rotated vector $\vec{x}^{\prime}$ with the help of a quantum circuit. In sync with the notation in other sections, the considerations is summarised as follows:
\begin{align}
    \vec{n} =  \frac{1}{\sqrt{6}}(2, 1, 1) \hspace{15mm} \vec{x} = \frac{1}{\sqrt{3}}(1 , 1, 1) \hspace{15mm} \theta_r = 90^{\circ} \label{eq:init_eq}
\end{align}
\subsection{Initialisation of Circuit to $\vec{x}$ on Bloch Sphere}
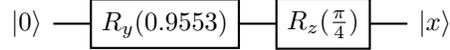
\begin{figure}[H]
    \centering
    \begin{tikzpicture}
        \node[scale=1]{
        \begin{quantikz}
            \lstick{\ket{0}} & \gate{R_y(0.9553)}
            & \gate{R_z(\frac{\pi}{4})} & \rstick{\ket{x}}\qw
        \end{quantikz}};
    \end{tikzpicture}
    \caption{Initialization Circuit as per equation \ref{eq:init_eq}}
    \label{fig:Eg_initCirc}
\end{figure}

We initialise the circuit parameters to represent $\vec{x}$ on the bloch sphere. Here the angles $\theta$ and $\phi$ described in Section \ref{sec:Bloch_sph} are 0.9553 and $\pi/4$ radians respectively. Thus the illustration in Figure \ref{fig:Eg_initCirc} depicts the circuit for initialization.
\subsection{Creating the Rotation Unitary}
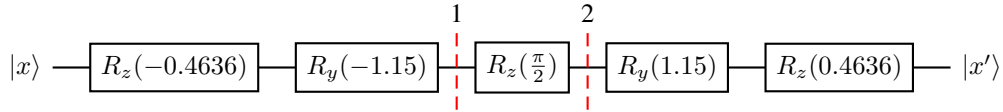
\begin{figure}[H]
    \centering
    \begin{tikzpicture}
        \node[scale=1]{
        \begin{quantikz}
        \lstick{\ket{x}} & \gate{R_z(-0.4636)} & \gate{R_y(-1.15)}\slice{1} & \gate{R_z(\frac{\pi}{2})}\slice{2} & \gate{R_y(1.15)} & \gate{R_z(0.4636)} & \rstick{\ket{x^\prime}} \qw
        \end{quantikz}};
    \end{tikzpicture}
    \caption{Illustration of transformation of initiated $\ket{x}$ to rotated $\ket{x^\prime}$ by the quantum circuit scheme}
    \label{fig:Eg_rotCirc}
\end{figure}

The $\theta$ and $\phi$ values for our chosen axis are 1.15 and 0.4636 radians respectively. Thus as discussed in earlier sections, to align this axis to the Z axis we need to rotate by $-\theta$ and $-\phi$. Now that our axis is aligned to the Z-axis, we rotate the vector by $\theta_r = 90^{\circ}$ as parameter to a $R_z$ gate to arrive at the rotated vector in this transformed vector space. To get the original rotated vector, we undo the earlier rotation by restoring the axis to earlier position. Fig. \ref{fig:Eg_rotCirc} depicts the circuit to carry out these three stages. 
    
It is also noteworthy that we can find the Euler-Rodrigues parameters solely from the unitary created here by using equation \eqref{eq:trace_properties}. The relevant values of $a$, $b$, $c$ and $d$ found for this particular case are 
\begin{align}
        a =  0.707; \hspace{10 mm }b =  0.577; \hspace{10 mm }c =  0.289; \hspace{10 mm }d =  0.289; \hspace{10 mm }
\end{align}
\subsection{Reconstructing the Rotated Vector}
\begin{figure}
\centering
\begin{subfigure}{0.4\textwidth}
  \hspace*{-2cm} 
  \centering
  \includegraphics[width=1.45\linewidth]{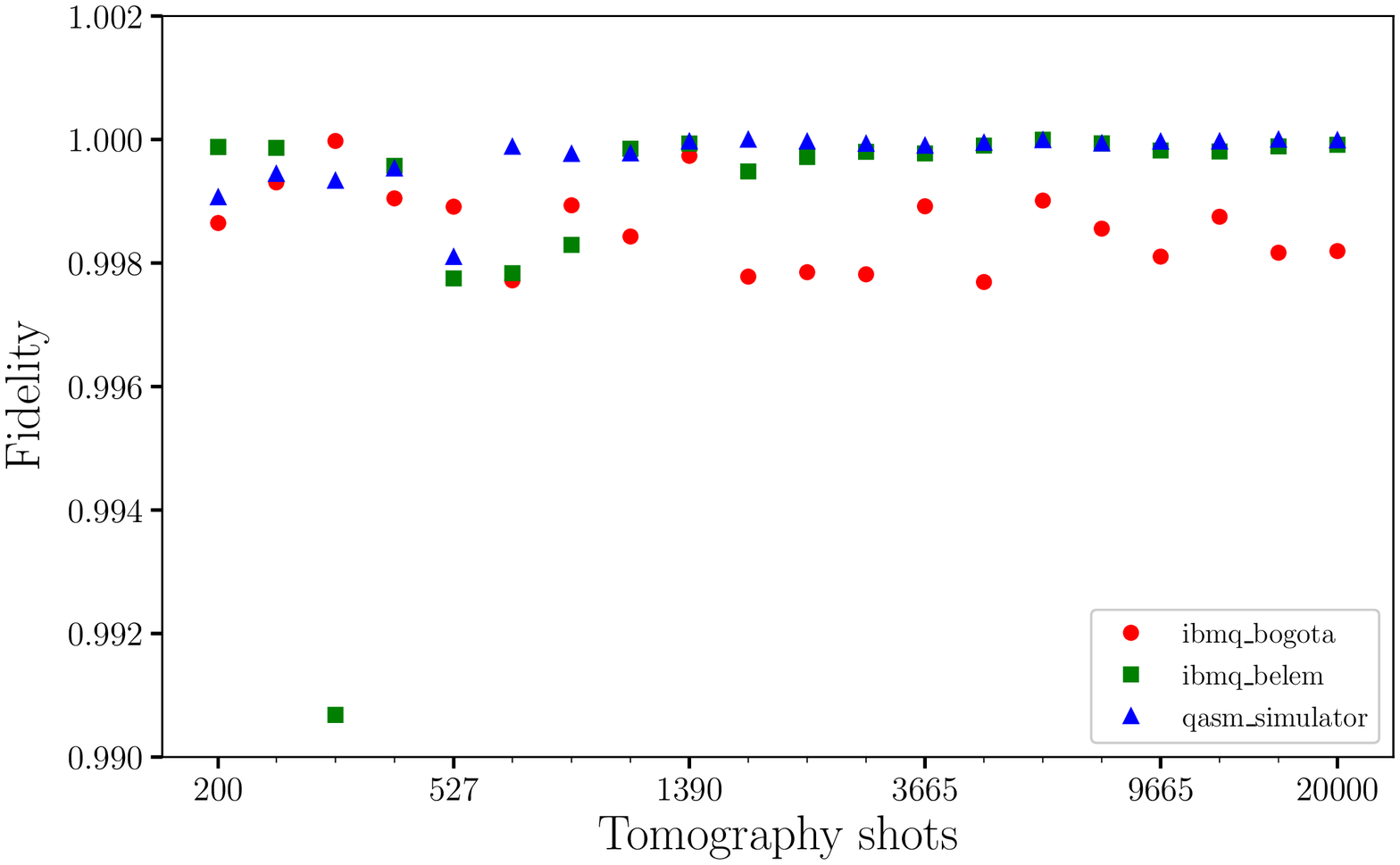}
  \label{fig:fidelity specific example}
  
\end{subfigure}%
\hspace{5mm}
\begin{subfigure}{0.4\textwidth}
  \hspace*{-0.5cm}
  \centering
  \includegraphics[width=1.45\linewidth]{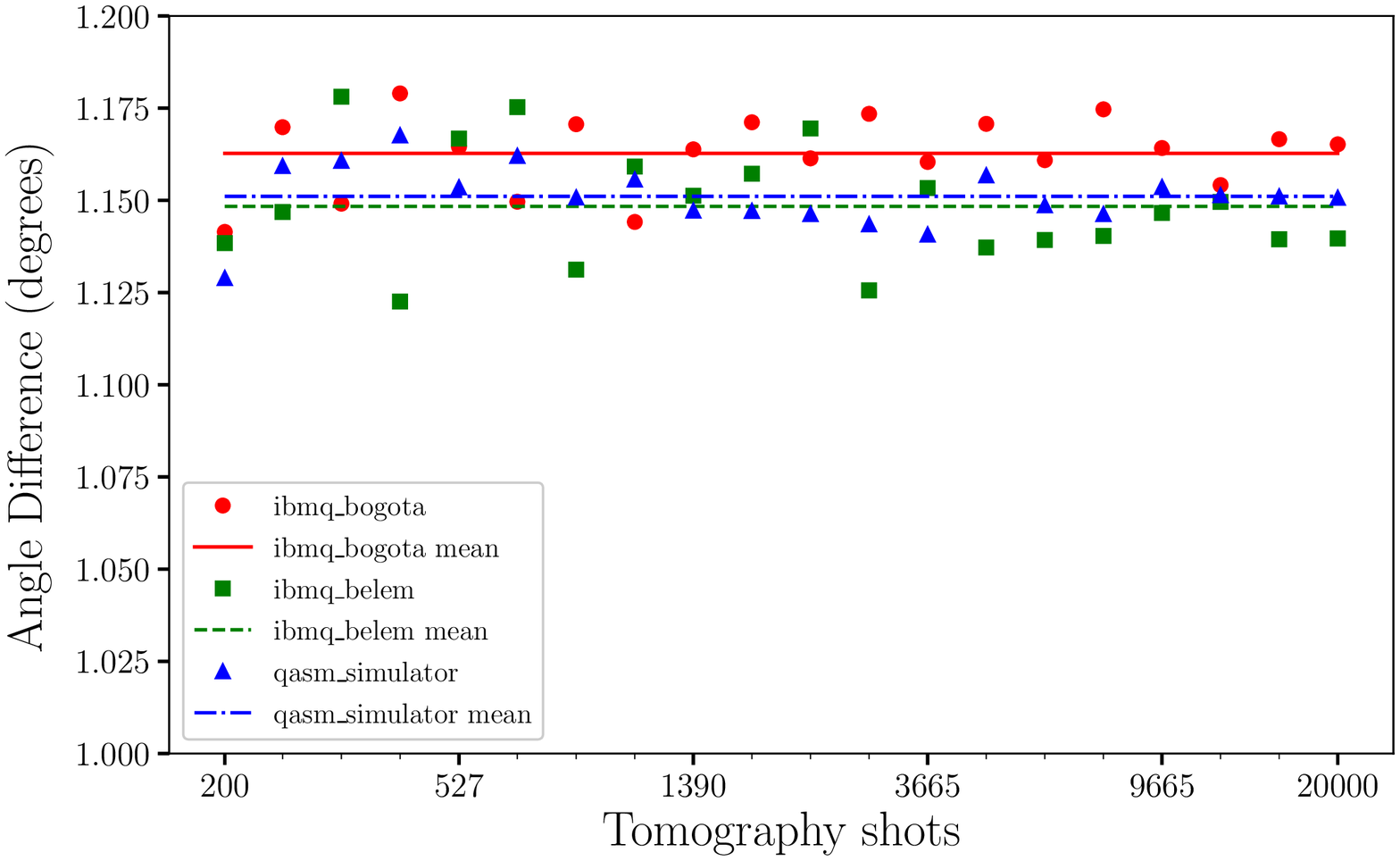}
  \label{fig:angle difference for our specifc example}
\end{subfigure}
\caption{Fidelity and Angle difference for our specific case with $\vec{n} =  \frac{1}{\sqrt{6}}(2, 1, 1)$ , $\vec{x} = \frac{1}{\sqrt{3}}(1 , 1, 1) $ \& $ \theta_r = 90^{\circ}$}
\label{fig:Fidelity and Angle difference for section}
\end{figure}

To reconstruct the rotated vector, we follow the method as described earlier. For running in real remote quantum devices, calibration matrix calculations were carried out in standard fashion. Finally results of all the run were fitted with either least square measure or by convex optimization. The Reconstructed vector error from the system can be intrepreted in terms of the fidelity of actual and expected density matrix. Also the angle between analytically solved and reconstructed vector can be a good metric of correctness. Figure \ref{fig:Fidelity and Angle difference for section} shows the relevent results for both real and simulator backends.


\section{Conclusion}

In this paper, we have drawn a relationship between the Euler-Rodrigues parameters for rotations in three-dimensional space and rotations on a single qubit by thinking of the gates in terms of their effect on the Bloch sphere. Then, we used process tomography in the resulting circuit to extract the Euler-Rodrigues parameters from the quantum circuit up to some desired accuracy which depends on the number of shots used in the tomography. This gives us an algorithm that takes as the rotation axis and angle of rotation in radians and outputs the Euler-Rodrigues parameters as desired. 

Using this characterization of the rotation parameters in a quantum circuit, we were able to rotate a vector by adding a vector initialization routine to our rotation single-qubit circuit. Then, we used state tomography (which requires less total circuits than its process counterpart) to extract the rotated vector. Using this as a basis, we proposed a multi-qubit rotation scheme to rotate $2^n$ vectors using $n+1$ qubits. This scheme allows us to encode the points that make up a rigid body and rotate it a given angle about some vector. A drawback to this approach is that the points need to be of unit length. 

Tests using Qiskit's \cite{Qiskit_2021} QASM simulator and three IBM quantum devices show that the techniques introduced in this paper yield good results when using the fidelity as a metric for the process tomography and the angle between the extracted vector and the exact rotated vector for state tomography. As expected, the larger the number of shots, the better the results. Fidelity $\geq 0.99$ starts showing around just three thousand shots for the real devices and before a thousand shots for the simulator. The angle difference is always less than a degree and remains quite stable for most number of shots, with IBMQ Belem showing the lowest difference at $\approx 0.72$ degrees.   

\bibliographystyle{unsrt}
\bibliography{references}

\end{document}